\documentclass[11pt]{article}
\usepackage{moriond,epsfig}

\def\be{\begin{equation}}
\def\ee{\end{equation}}
\def\bea{\begin{eqnarray}}
\def\eea{\end{eqnarray}}

\newcommand{\smto} {{\ifmmode \kern0.08em\to\kern0.08em%
                     \else $\kern0.08em\to\kern0.08em$ \fi}}
\newcommand{\btol} {{\ifmmode \mathrm{b}\!\to\!l%
                     \else $\mathrm{b}\!\to\!l$ \fi}}
\newcommand{\ctol} {{\ifmmode \mathrm{c}\!\to\!l%
                     \else $\mathrm{c}\!\to\!l$ \fi}}
\newcommand{\bctol}{{\ifmmode \mathrm{b}\!\to\!\mathrm{c}\!\to\!l
                     \else $\mathrm{b}\!\to\!\mathrm{c}\!\to\!l$\fi}}
\newcommand{\bctolidx}
       {{\ifmmode%
         b\kern-0.03em\to\kern-0.05emc\kern-0.03em\to\kern-0.05eml%
         \else
         $b\kern-0.03em\to\kern-0.05emc\kern-0.03em\to\kern-0.05eml$%
         \fi}}
\newcommand{\Bar}[1]{{\,\overline{\!#1}}}
\newcommand{\sinefflep}   {{\ifmmode  \sin^2\!\theta%
                          _\mathrm{eff}^\mathrm{\kern0.10emlept}
                          \else    $\sin^2\!\theta%
                          _\mathrm{eff}^\mathrm{\kern0.10emlept}$\fi}}
\newcommand{\ga}[1]{\ifmmode g_{A#1} \else $g_{a#1}$ \fi}
\newcommand{\gv}[1]{\ifmmode g_{V\!#1} \else $g_{v#1}$ \fi}
\newcommand{\alr}{\ensuremath{A_{\mathrm{LR}}}}
\newcommand{\afb}{\ensuremath{A_{\mathrm{FB}}}}
\newcommand{\afbqborn}{\ensuremath{A_{\mathrm{FB}}^{\mathrm{q},\,0}}}
\newcommand{\afbbborn}{\ensuremath{A_{\mathrm{FB}}^{\mathrm{b},\,0}}}
\newcommand{\afbcborn}{\ensuremath{A_{\mathrm{FB}}^{\mathrm{c},\,0}}}
\newcommand{\afbbwhat}[1]{\ensuremath{A_{\mathrm{FB}}^{\mathrm{b},\,#1}}}
\newcommand{\afbxwhat}[2]{\ensuremath{A_{\mathrm{FB}}^{#1,\,#2}}}
\newcommand{\afbc}{\ensuremath{A_{\mathrm{FB}}^\mathrm{c}}}
\newcommand{\afbb}{\ensuremath{A_{\mathrm{FB}}^\mathrm{b}}}
\newcommand{\afbf}{\ensuremath{A_{\mathrm{FB}}^f}}

\newcommand{\purity}{\ensuremath{\mathcal{P}}}
\newcommand{\fontexperiment}[1]{{\small #1}}
\newcommand{\cAleph}{\fontexperiment{ALEPH}}
\newcommand{\cDelphi}{\fontexperiment{DELPHI}}
\newcommand{\cLdry}{\fontexperiment{L3}}
\newcommand{\cOpal}{\fontexperiment{OPAL}}
\begin{document}
\vspace*{4cm}
\title{HEAVY FLAVOUR ELECTROWEAK PHYSICS REVIEW}

\author{W. LIEBIG }

\address{Bergische Universit\"at-GH, Gau\ss{}str. 20, D-42097
Wuppertal, Germany\\
\texttt{Wolfgang.Liebig@cern.ch}}

\maketitle\abstracts{
The four LEP experiments and the SLD detector
have measured the Z partial widths in the processes
$e^+e^-\smto{}\mathrm{Z}\smto{}\mathrm{b}\overline{\mathrm{b}}$ and  
$e^+e^-\smto{}\mathrm{Z}\smto{}\mathrm{c}\overline{\mathrm{c}}$ and
the corresponding forward-backward asymmetries 
at centre-of-mass energies close to $m_\mathrm{Z}$.
The results yield a very precise determination of
the effective vector and axial-vector coupling constants
and of the underlying electroweak mixing angle $\sinefflep$,
probing the Standard Model prediction for the electroweak
radiative corrections.
Of special interest is hereby a difference at the level of three
standard deviations in the mixing angle results from lepton
production and those obtained from the forward-backward asymmetry
in b quark production.
The b quark asymmetry measurements, some of which are still in the
process of being finalised, are therefore discussed in detail.
}

%================================================
\section{Theoretical Motivation}
%================================================
\label{s:theory}%

The forward-backward asymmetries constrain the product of the
asymmetry parameters for initial electrons and the final b and c
quarks \cite{lep1yb},
\begin{equation}
  \afbqborn = {3\over 4}{\mathcal A}_{e}{\mathcal A}_\mathrm{q}
%            = {3\over 4} \frac{2\ga{e}
%    \gv{e}}{\ga{e}^2+\gv{e}^2}
%    \frac{2\ga{q} \gv{q}}{\ga{q}^2+\gv{q}^2}~.
  \label{eqn:afbcoupl}
\end{equation}
%
% This product is determined mainly by ${\mathcal A}_e$
% because the Standard Model values of the electric charge
% and the iso-spin for quarks lead to ${\mathcal A}_q$
% values for b and c quarks close to 1.
The Standard Model values of the electric charge
and the iso-spin for quarks lead to ${\mathcal A}_\mathrm{q}$
values for b and c quarks close to 1, hence it is ${\mathcal A}_e$
that $\afbqborn$ is mainly sensitive to.
${\mathcal A}_e$ is a function of the ratio of the effective vector
and axial-vector neutral current couplings, $\gv{l}/\ga{l}$,
and thus of the effective electroweak mixing angle $\sinefflep$.
As a consequence, the results from the forward-backward asymmetry
of b and c quarks can be combined with other lepton measurements
to determine $\sinefflep$ with high precision, and to put constraints
on the Higgs boson and top quark mass that enter the effective mixing
angle via higher order electroweak corrections.
On the other hand, the comparison between the heavy flavour
asymmetries and other asymmetry measurements probe the internal
consistency of the Standard Model.

In contrast, the measurements of the Z partial decay widths
determine the sum of the squares of the coupling constants.
In combination with the asymmetry results it is hence possible
to disentangle $\ga{q}$ and $\gv{q}$.
%
%================================================
\section{Heavy Flavour Electroweak Measurements}
%================================================
The LEP and SLC results for $R_{\mathrm b}$ and $R_{\mathrm c}$ are
final since 2000 \cite{lepew-rbrc}.
The tools to select b and c events were developed in the framework of
these analyses and have been transferred to the asymmetry
measurements, where the main experimental task is then to
distinguish between the flight directions of the quark and the
antiquark.
This is achieved by reconstructing the quark charge in one or both of
the hemispheres defined by the thrust axis. Two different and largely
independent reconstruction methods are applied: one directly yields the
clear charge information in case of a weak B decay into a lepton and
measures $\afbb$ and $\afbc$ simultaneously by means of b-c separation 
variables. The other approach analyses inclusively all decay types,
requiring a flavour-pure data set obtained from b-tagging.
Before the asymmetry results are discussed, the lepton and the
inclusive method and their sophisticated analysis tools will be
briefly presented.
%
%==============================================================
\section[$\afbb$ and $\afbc$ with Leptons]
{\boldmath$\afbb$ and $\afbc$ with Leptons\unboldmath}
%==============================================================
All LEP experiments provide measurements of 
$\afbb$ and $\afbc$ using prompt leptons that are produced with
large $p$ and $p_\perp$ in weak decays of b and c hadrons\,%
\cite{alefSL,dlfiSL,ldrySL,opalSL}.
The $p$, $p_\perp$ of identified electrons and muons are
the principal means of distinguishing between
\btol, \ctol, \bctol{} decays and background.
For the purpose of identifying the decay type, new techniques
use either neural networks or likelihood ratios to
combine the $p$, $p_\perp$ measurement with additional lifetime and
kinematic information from the rest of the b- or c-jet.
Depending on the experiment, also the jet charge from the
opposite hemisphere is included.
This has improved the sensitivity to B-$\overline{\mathrm{B}}$
mixing and allowed to measure the b asymmetry also via wrong-sign
\bctol{} decays.
$\afbb$ and $\afbc$ and possibly the mixing parameter $\chi$ are
then extracted from a fit in bins of the decay type identifiers
and of $\cos{\theta}$.
The c asymmetry obtained from this method,
$\afbcborn\,\mbox{(leptons)} = 0.0700\pm0.0034$,
is consistent with the result from exclusively reconstructed D meson
decays\,\cite{alefD*,dlfiD*,opalD*},
$\afbcborn\,\mbox{(excl. D}{}^*) = 0.0711\pm0.0057$.
%
%
%==============================================================
\section[Inclusive Jet Charge and New Neural Net Analyses for $\afbb$]
        {Inclusive Jet Charge and New Neural Net Analyses 
            for \boldmath$\afbb$\unboldmath}
%==============================================================

\begin{figure}
\begin{minipage}{0.44\linewidth}
  \psfig{figure=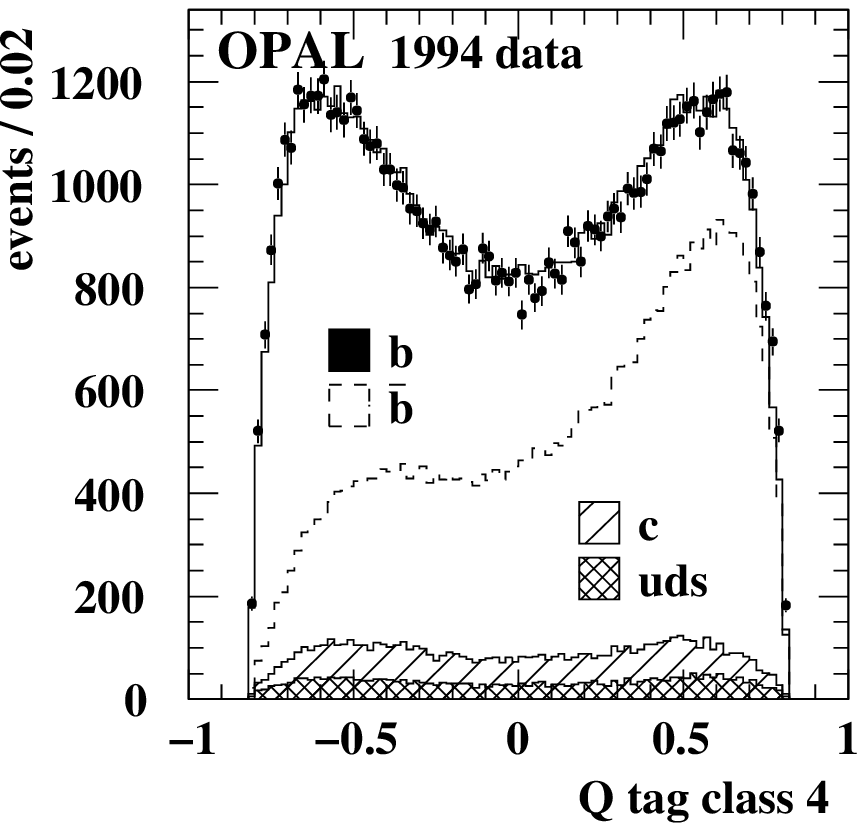,
         width=1.0\linewidth,bb= 0 0 249 236}
\end{minipage}\qquad
\begin{minipage}{0.52\linewidth}
  \psfig{figure=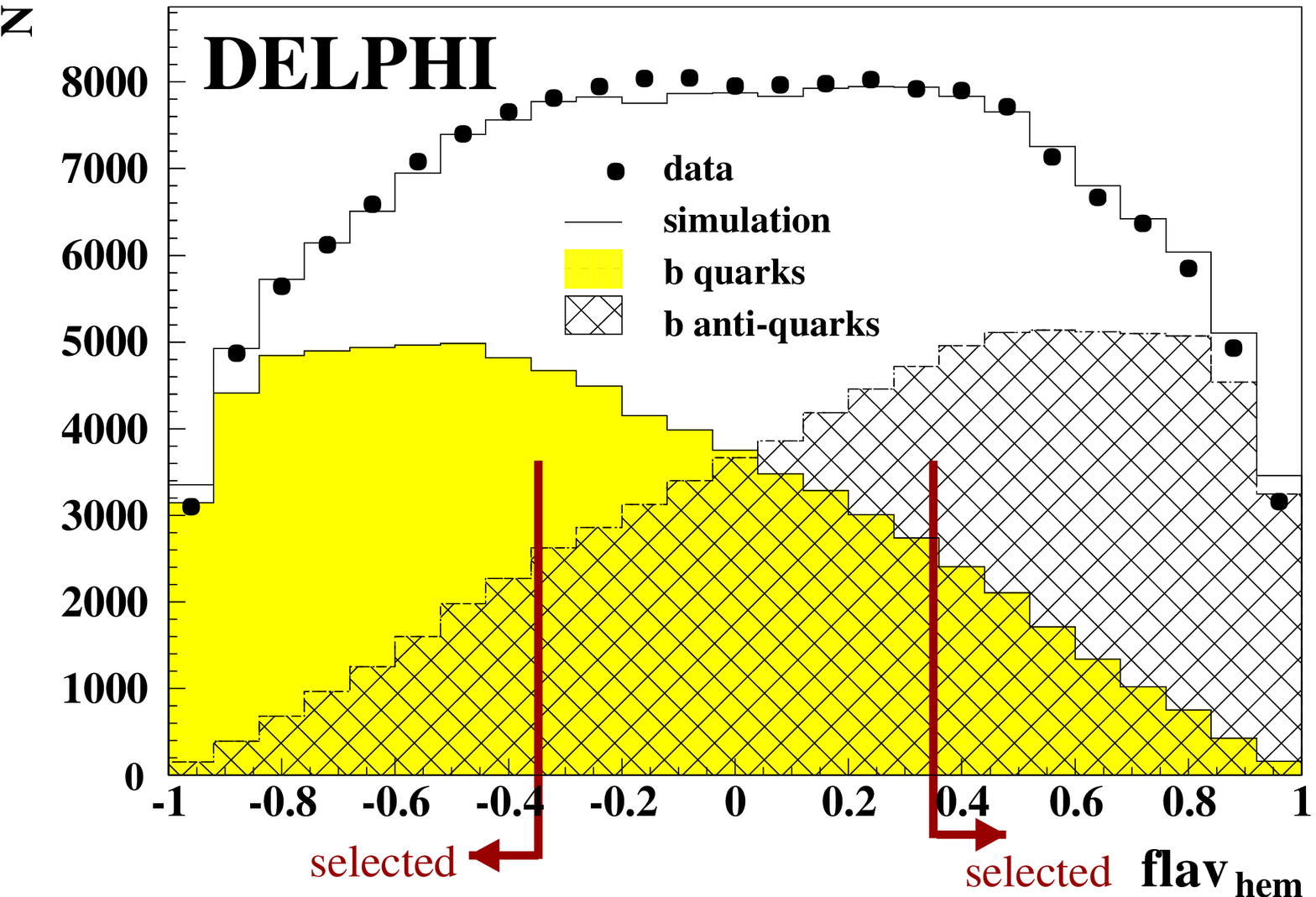,
         width=1.0\linewidth,bb=  0 0 569 388}
\end{minipage}
\caption{The output distribution of the \cOpal{} and \cDelphi{}
         charge tagging neural networks for the data
         of 1994 in comparison with simulation.
\label{fig:nnoutputs}}
\end{figure}
All four LEP experiments exploit also an inclusive approach to measure 
$\afbb$ from a large size b-tagged sample with usually very low
backgrounds \cite{dlfiJC,ldryJC}.
The jet charge that is used in these analyses is correlated to the
primary quark, however the charge dilution by fragmentation and
subsequent decays limits the sensitivity to $\afbb$.
Therefore \cAleph{}, \cDelphi{} and \cOpal{} have recently
performed \cite{alefNN,dlfiNN,opalNN}
improved measurements that make use of various pieces of
information on b-jets combined in a neural network
whose output is correlated to the primary b quark charge.
The information comprises not only the jet charge
% at various weighting powers $\kappa$,
but also the secondary vertex charge that provides a reliable
charge reconstruction in case of long-lived $\mathrm{B}^\pm$ mesons.
Additional precision is gained by including identified particles,
be it in the
$\mathrm{b}\!\to\!\mathrm{c}\!\to\!\mathrm{s}\!\to\!\mathrm{K}^-$
decay chain, or via more inclusive $\mathrm{B}_x$ charge tags that
weight the charge at the secondary and primary vertex depending on the
B hadron hypothesis.
The output of such a network 
is displayed for the \cOpal{} and \cDelphi{} data of 1994 in
Fig.~\ref{fig:nnoutputs}, showing a very pure $\overline{\mathrm{b}}$
(b) identification at large positive (negative) values.
%
%--------------------------------------------
% \subsection{The analyses}
%--------------------------------------------
Based on its charge output $Q$, $\afbb$ is extracted from the average
charge flow%
\vspace*{-0.3ex}
\begin{equation}
  \left<Q_{\mathrm{F}} - Q_{\mathrm{B}}\right>
       = \!\sum_{f}{\purity_f(\theta)%
                    \delta_f(\theta)\cdot\afbf\cdot%
               \frac{8}{3}\frac{\cos{\theta}}{(1+\cos^2{\theta})}
                 }\quad,
\label{eqn:alnn-afb}\vspace*{-0.3ex}
\end{equation}
using a very comprehensive hemisphere tag (\cAleph)
or separate classes of differently composed tags (\cOpal)
to build $Q_{\mathrm{F}}-Q_{\mathrm{B}}$.
\cDelphi{} uses the counting method%
\vspace*{-0.3ex}
\begin{equation}
  \!\begin{array}{l}\displaystyle%
      \frac{N_{\mathrm{F}}-N_{\mathrm{B}}}
           {N_{\mathrm{F}}+N_{\mathrm{B}}} =
       \displaystyle\sum_{f}{\purity_f(\theta)(2w_f(\theta)-1)%
                \cdot\afbf\cdot%\\[0.1ex]
               \frac{8}{3}\frac{\cos{\theta}}{(1+\cos^2{\theta})}}
  \end{array}%\vspace*{-1ex}
  \label{eqn:dlnn-afb}\vspace*{-0.3ex}
\end{equation}
with a cut-based selection of hemispheres.
The measurements work on b-enriched samples with b purities of about
$90\,\%$, with the flavour fractions $\purity_f$ extracted mainly from
the data.
The calibration of the charge correlation to the initial b quark
($\delta_\mathrm{b}$ in Eq.~\ref{eqn:alnn-afb} and
$(2w_\mathrm{b}-1)$ in Eq.~\ref{eqn:dlnn-afb}) 
is very important for being independent from B-physics
modelling in the simulation.
Therefore \cAleph{} and \cOpal{} calibrate the
charge separation $\delta_\mathrm{b}$ from the widths of the
charge flow and total charge distributions by measuring
\begin{equation}
  \Bar{\delta}^2 = \sigma^2(Q_{\mathrm{F}} - Q_{\mathrm{B}})
                    - \sigma^2(Q_{\mathrm{F}} + Q_{\mathrm{B}})
     \approx\sum_f{\purity_f\delta^2_f}\quad.
  \label{eqn:alnn-calib}
\end{equation}
\cDelphi{} calibrates the probability $w_\mathrm{b}$ to identify 
the quark charge correctly from the numbers of like and
unlike sign charged hemisphere pairs.
% \begin{equation}
%   \begin{array}{rcl}
%        N\Bez{unlike} &\approx& w_b^2 + (1-w_b)^2\\
%        N\Bez{like}   &\approx& 2\cdot{}w_b\cdot{}(1-w_b)
%     \end{array}
%   \label{dlnn-calib}
% \end{equation}
%
\begin{figure}
\begin{minipage}{0.44\linewidth}
  \psfig{figure=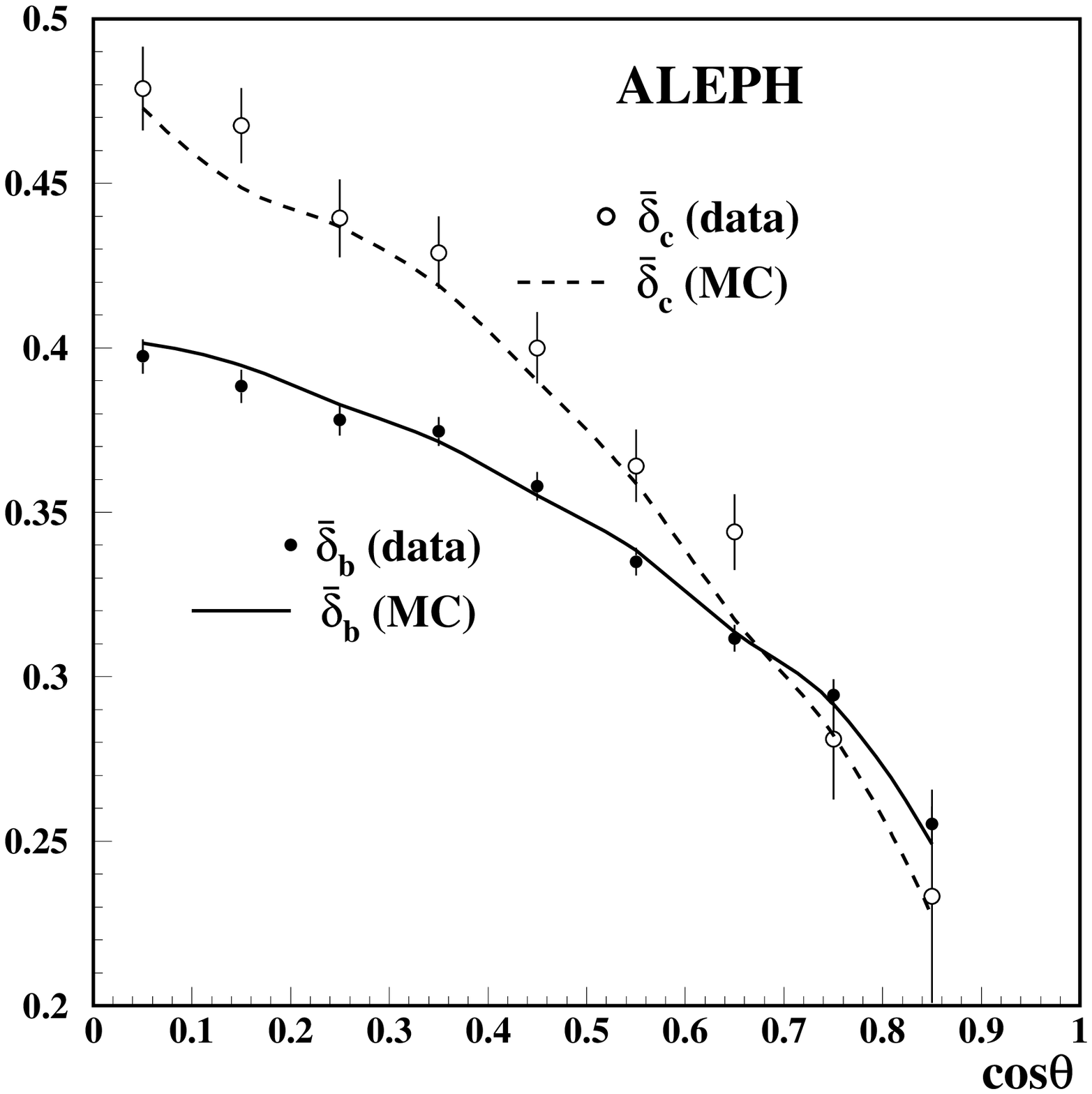,
         width=1.0\linewidth,bb= 16 19 514 516}
\end{minipage}\qquad
\begin{minipage}{0.52\linewidth}
  \psfig{figure=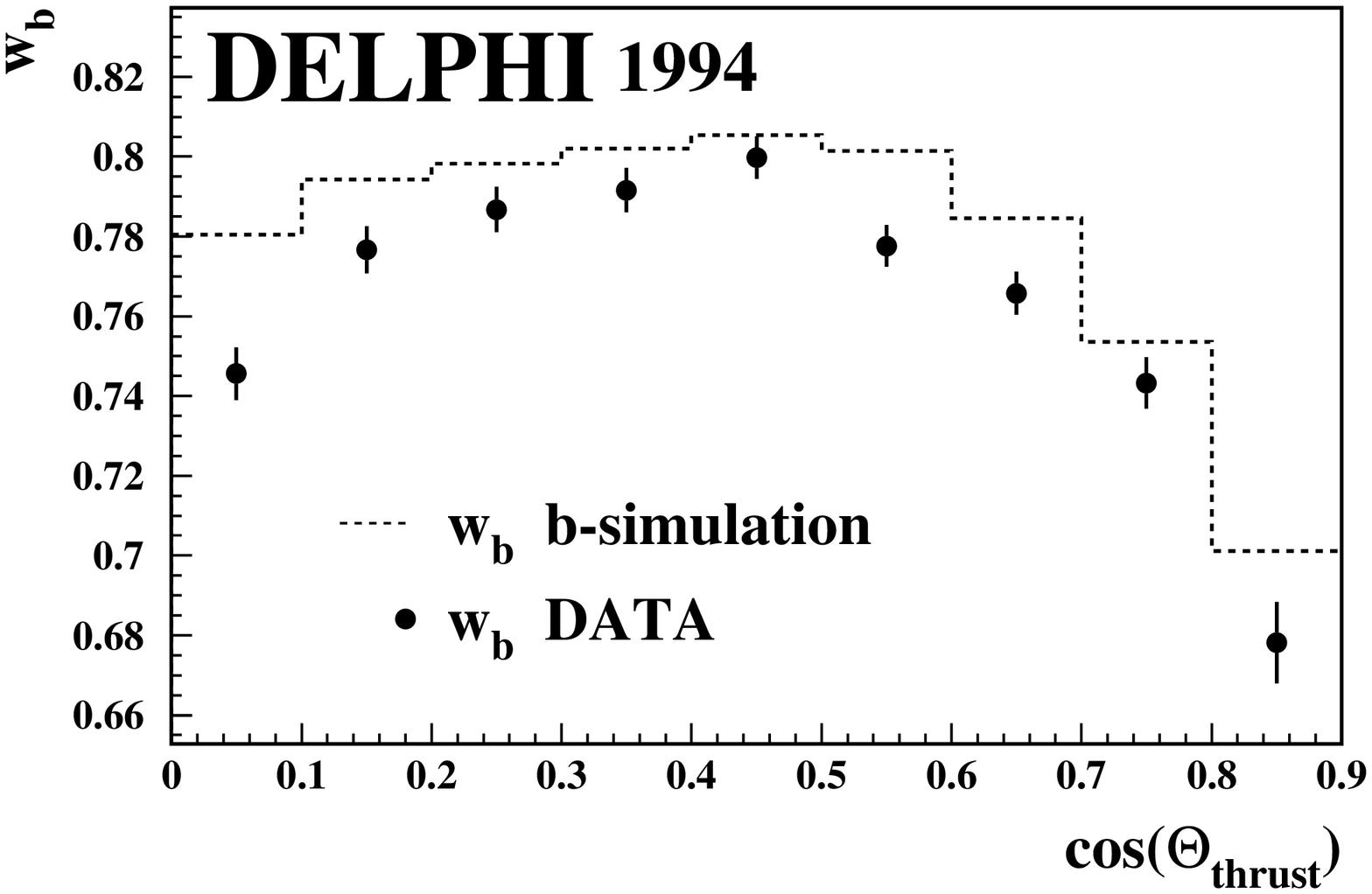,
         width=1.0\linewidth,bb= 16 317 568 682}
\end{minipage}
\caption{The sensitivities to the b asymmetry in the \cAleph{} and
         \cDelphi{} inclusive measurments. The direct measurement
         on the data is compared to the prediction from the simulation.
\label{fig:calibration}}
\end{figure}%
The calibrated $\delta_f$ and $w_\mathrm{b}$
are shown in Fig.~\ref{fig:calibration} for the \cAleph{} and
\cDelphi{} analyses.
$\afbb$ is then extracted from a fit to the differential
asymmetry.
The \cDelphi{} result from the inclusive b asymmetry measurement\,%
\cite{dlfiNN} has been updated, including the LEP\,1 off-peak
data as well as the LEP\,2 Z runs.
Key quantities are now calibrated also for the charm background:
a double tagging method corrects the charm fraction while
reconstructed $\mathrm{D}^*$ mesons in the opposite hemisphere
are used to correct the charm charge sensitivity $(2w_\mathrm{c}-1)$.
The new result presented at this
conference is $\afbbborn = 0.0978\pm0.0030(stat.)\pm0.0014(syst.)$
and will be included in the LEP combined b asymmetry in summer 2003.
%
%--------------------------------------------
\subsection{Modelling Dependence and Checks}
%--------------------------------------------
%
The remaining corrections that have to be taken from simulation
are the individual sensitivity to the hemisphere correlations
and the QCD correction.
The hemisphere correlations arise mainly from charge conservation in
the event and gluon radiation. They amount to $5-9\,\%$ in
linear correction factors to $w_b$ and $\delta_b$.
Extensive checks have been performed by the experiments,
covering the thrust dependence, monitoring the response of the
correlation to selected network inputs and comparing to observables on
the real data that are sensitive to the hemisphere correlation.
%
% like OPAL's
%           \hat{C} = \frac{\left<|X_F|\cdot|X_B|\right>
%                - \left<|X_F|\right>\cdot\left<|X_B|\right>}
%                {\sigma_{X_F}\sigma_{X_B}}
Their results are included in the systematic uncertainties.

The fitted asymmetry values have to be corrected for gluon radiation
from the primary quark pair and for approximating their direction of
flight by the thrust axis. This QCD correction has been calculated in
${\mathcal O}(\alpha_s^2)$ to be
$\afbbwhat{\mathrm{fit}} = (1 - 0.0354\pm0.0063)\cdot\afbb$,
which is more than double the statistical uncertainty of the LEP
combined asymmetry result.
However the selection of high purity b events biases against events
with hard gluon radiation, 
so that only about 1/4 or less of the full effect is observed.
Therefore the residual sensitivity to the QCD effects is computed
individually in each analysis and the asymmetries are corrected
to the quark level.
Studies of the thrust cut dependence in the improved jet-charge
analyses revealed that the correction for hemisphere charge
correlations in the determination of the charge tagging power
directly from the data also absorbs part of the QCD correction.
Hence both residual effect and hemisphere correlations
are taken into account for a safe estimation of the common systematic
uncertainty due to the QCD correction.
%
%---------------------------
% --  AFB interpretation  --
%---------------------------
\section{Asymmetry Results in the Light of the Standard Model}
\begin{table}[b]
  \begin{center}
  \begin{tabular}{|l|c|}\hline\rule{0pt}{2.6ex}
    ``leptons'' -- $\afbxwhat{l}{0}$,
      ${\mathcal A}_{l}(P_\tau)$,
      $\alr$ & $0.23113 \pm 0.00021$\\[0.3ex] \hline\rule{0pt}{2.6ex}
    ``hadrons'' -- $\afbbborn$, $\afbcborn$, $<Q_\mathrm{FB}>$
             & $0.23217 \pm 0.00029$\\[0.3ex] \hline
  \end{tabular}
  \end{center}
  \caption{Results for $\sinefflep$ from different groups of
           asymmetry measurements}\vspace*{-1ex}%
  \label{t:sinthw}
\end{table}%
\begin{figure}
\begin{minipage}{0.54\linewidth}\hspace*{-0.7em}
  \psfig{figure=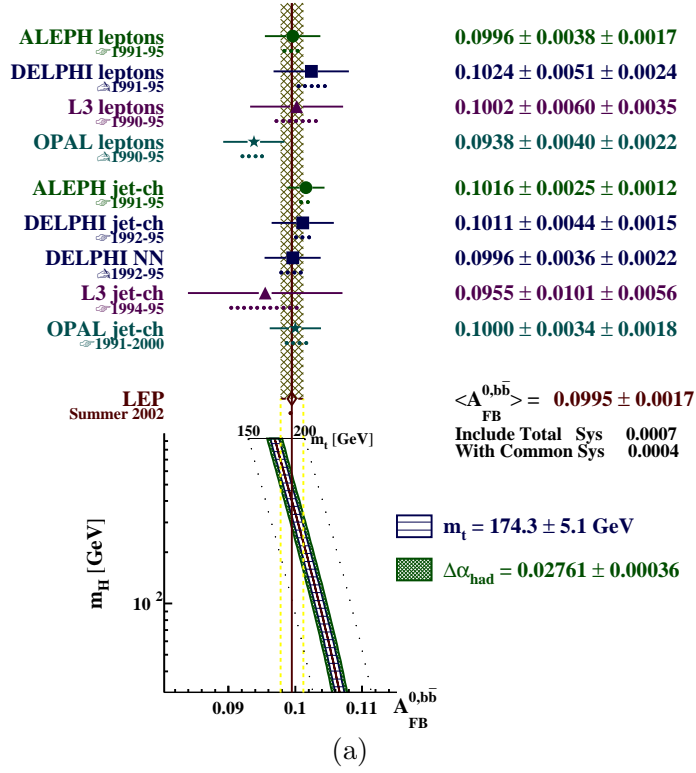,
         width=1.05\linewidth,bb=  28 29 560 595}
\end{minipage}\qquad
\begin{minipage}{0.44\linewidth}\hspace*{-1em}
  \psfig{figure=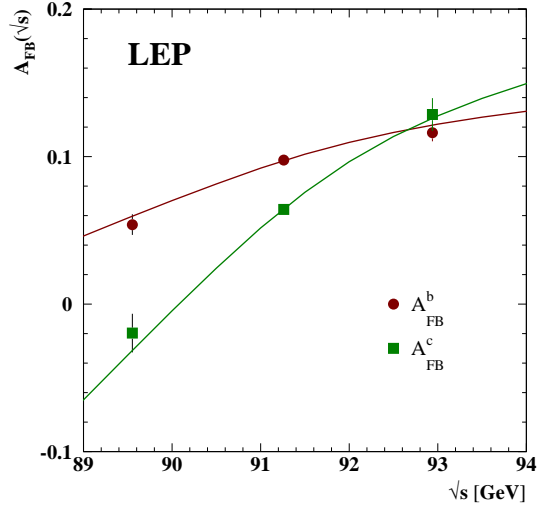,
         width=1.15\linewidth,bb= 0 0 567 567}
\end{minipage}\\[-0.6ex]
\begin{minipage}{0.54\linewidth}\hspace*{-0.7em}
\begin{center}
  (a)
\end{center}
\end{minipage}\qquad
\begin{minipage}{0.44\linewidth}\hspace*{-1em}
\begin{center}
  (b)
\end{center}
\end{minipage}
\caption{The LEP results for $\afbbborn$.
         (a) The average takes into account correlated errors and
         off-peak measurements.
         (b) The results for the b and c asymmetry vs. centre-of-mass
         energy.\vspace{-2.5ex}
\label{fig:lepw03-afbb0}}
\end{figure}
The results for the b quark pole asymmetry $\afbbborn$ from
the four lepton and five inclusive analyses are shown in
Fig.~\ref{fig:lepw03-afbb0}a.
The average statistical correlation between the two types of analyses
is $\sim25\%$. The results represent the status of winter
2003\,\cite{lepew-w03} and are very consistent with each other.
By taking into account correlations and common systematic errors as well
as off-peak measurements (Fig.~\ref{fig:lepw03-afbb0}b) in cases
where available one obtains the LEP combined result of
\begin{equation}
  <\afbbborn> = 0.0995\pm0.0017\quad.
  \label{eqn:afbb0result}
\end{equation}
The correlated systematic error arises from mainly physics like 
the QCD correction and light quark fragmentation.
Its value of $0.0004$ is quoted in Fig.~\ref{fig:lepw03-afbb0}a
and turns out to be very small.
According to Section~\ref{s:theory}, the measurement of $\afbbborn$
can be used together with direct lepton $\afb$ measurements and 
$\alr$ from SLD\,\cite{bib:sldalr}\ to determine $\sinefflep$.
\begin{figure}\vspace*{2ex}%
\begin{minipage}{0.44\linewidth}\hspace*{-1.5em}
  \psfig{figure=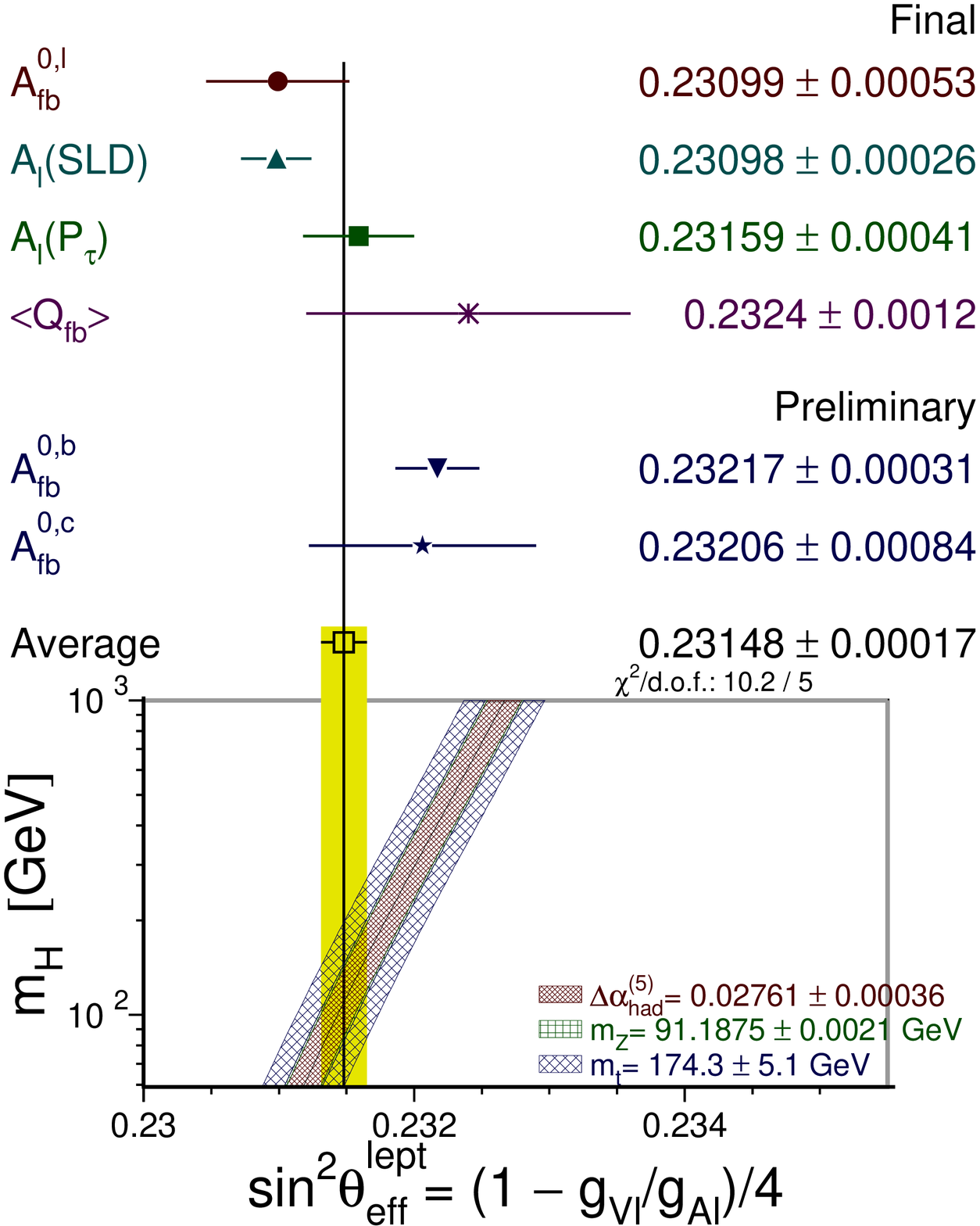,
         width=1.02\linewidth,bb=  5 1 561 699}
\end{minipage}\qquad\
\begin{minipage}{0.52\linewidth}
  \psfig{figure=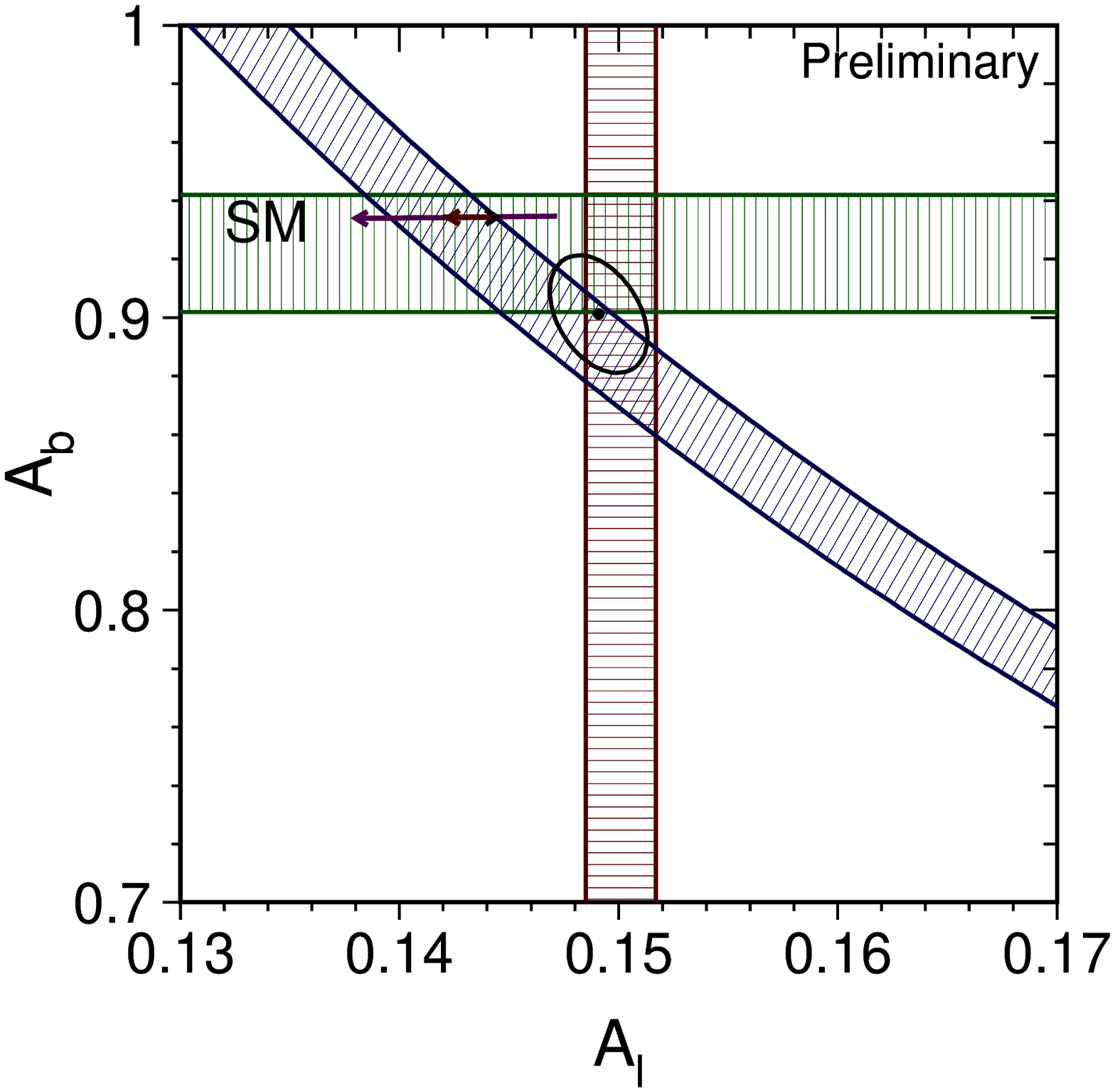,
         width=1.0\linewidth,bb= 0 0 567 624}
\end{minipage}%
\\[-0.6ex]
\begin{minipage}{0.44\linewidth}
\begin{center}
  (a)
\end{center}
\end{minipage}\qquad
\begin{minipage}{0.52\linewidth}
\begin{center}
  (b)
\end{center}
\end{minipage}\vspace*{-0.5ex}
\caption{Interpretation of the $\afbb$ results in terms of the
         asymmetry parameters and the leptonic effective
         electroweak mixing angle.
         The right plot shows $\pm1\sigma$ bands in the
         $({\mathcal A}_l,{\mathcal A}_\mathrm{b})$ plane for
         ${\mathcal A}_l$ (from the $\alr$, $\afbxwhat{l}{0}$, and
         $\tau$ polarisation measurements),
         for ${\mathcal A}_\mathrm{b}$
         (from the SLD polarised $\afbb$) and for
         $\afbbborn=(3/4){\mathcal A}_l{\mathcal A}_\mathrm{b}$
         (diagonal band).
         The small dependence of the Standard Model prediction on
         ${\mathcal A}_\mathrm{b}$ allows a direct comparison
         in terms of $\sinefflep$, as shown in the Figure (b) on the
         left.
\label{fig:lepw03-sinthw}}
\end{figure}
This combination is illustrated in Fig.~\ref{fig:lepw03-sinthw}a
and shows that the lepton and heavy quark results 
are not very consistent with each other; a situation that exists
since the beginning of asymmetry measurements at LEP and SLC.
The averages of the two groups of measurements are listed in
Table~\ref{t:sinthw}.
All six measurements give a combined $\sinefflep$ result of
$0.23148\pm0.00017$ with a fit probability of $7\%$.

The electroweak fit itself gives no clear hint of a problem in the
Standard Model.
For example Fig.~\ref{fig:lepw03-sinthw}b shows that 
LEP and SLC are not inconsistent if both ${\mathcal A}_\mathrm{b}$
and ${\mathcal A}_l$ are free parameters, there is a region of common
overlap. Although the overlap region agrees poorly with the Standard
Model expectation, a single result like $\afbbborn$ from LEP can still
be well compatible.
Via higher order corrections to \sinefflep, $\afbbborn$ is
sensitive to $m_\mathrm{h}$. Interestingly it is the only quantity that
prefers a high Higgs boson mass, thus making the electroweak fit more
compatible with the direct exclusion limit on $m_\mathrm{h}$
\cite{lepew-w03,afbdoubt2}.

%
%--------------------
% --  Conclusion   --
%--------------------
\section{Concluding Remarks}
Heavy flavour foward-backward asymmetries have been measured at LEP
with various powerful and sophisticated techniques
that have set great store on model-independence and self-calibration.
The precise results are self-consistent and have a small common
systematic error, suggesting that the difference to the combined
mixing angle based on leptonic results is rather of statistical
nature, if not a hint for new physics.
The interpretation of the heavy flavour electroweak results
is unchanged since summer 2002, but a conclusive word on asymmetries
from LEP is expected in summer 2003 when the final results from
DELPHI and OPAL are included in the combination.

\end{document}